\RequirePackage{ifpdf}
\ifpdf 
\documentclass[pdftex]{sigma}
\else
\documentclass{sigma}
\fi

\numberwithin{equation}{section}

\begin{document}

\allowdisplaybreaks

\renewcommand{\thefootnote}{$\star$}

\renewcommand{\PaperNumber}{042}

\FirstPageHeading

\ShortArticleName{Potentials Unbounded Below}

\ArticleName{Potentials Unbounded Below\footnote{This
paper is a contribution to the Proceedings of the Workshop ``Supersymmetric Quantum Mechanics and Spectral Design'' (July 18--30, 2010, Benasque, Spain). The full collection
is available at
\href{http://www.emis.de/journals/SIGMA/SUSYQM2010.html}{http://www.emis.de/journals/SIGMA/SUSYQM2010.html}}}

\Author{Thomas CURTRIGHT~$^{\dag\ddag}$}

\AuthorNameForHeading{T.~Curtright}

\Address{$^\dag$~CERN, CH-1211 Geneva 23, Switzerland}
\Address{$^\ddag$~Department of Physics, University of Miami, Coral Gables, FL 33124-8046, USA}
\EmailD{\href{mailto:curtright@physics.miami.edu}{curtright@physics.miami.edu}}
\URLaddressD{\url{http://curtright.com/}}

\ArticleDates{Received December 21, 2010, in f\/inal form March 27, 2011;  Published online April 26, 2011}

\Abstract{Continuous interpolates are described for classical dynamical systems def\/ined
by discrete time-steps.   Functional conjugation methods play a central role
in obtaining the interpolations.  The interpolates correspond to particle
motion in an underlying potential,~$V$.   Typically, $V$ has no lower bound
and can exhibit switchbacks wherein $V$ changes form when turning points are
encountered by the particle.   The Beverton--Holt and Skellam models of
population dynamics, and particular cases of the logistic map are used to
illustrate these features.}

\Keywords{classical dynamical systems; functional conjugation methods; Beverton--Holt model; Skellam model}

\Classification{37C99; 37D45; 37E05; 37J05; 37M99; 39B22}

\renewcommand{\thefootnote}{\arabic{footnote}}
\setcounter{footnote}{0}

\section{Introduction}

The main purpose of this paper is to provide a few new results obtained from
the application of functional conjugation methods~-- Schr\"{o}der theory~\cite{s}~-- to problems in dynamics.  Specif\/ically, the interdisciplinary
connections of the Goldstone potential~\cite{g} to the population dynamics
model of Beverton and Holt~\cite{b}, and of qualitatively similar potentials
to the mathematically more intricate population model of Skellam \cite{sk}, as
described below in Section~\ref{section3}, are original, so far as I~am aware\footnote{This
statement is reinforced by the fact that my colleagues in mathematical ecology
and biology were unaware of any such connections.}. For an illuminating
discussion of these and other population models, and an introduction to the
biology literature, see~\cite{gk}.

Moreover, while the discrete logistic map, $x\mapsto sx(  1-x)  $,
has been particularly well-studied, especially since the work of Feigenbaum~\cite{f}, my discussion of the potentials that underlie the map is based on a
new investigation using Mathematica code to construct series solutions to
several hundred orders, and then to functionally continue those series.  The
remarkable and interesting features of the model's phase-space trajectories
are again novel, and~-- at least for the $s=3$ case presented in Section~\ref{section4}~--
displayed here for the f\/irst time, so far as I know.   Although the potentials
for the logistic map are given by convergent series for $s\neq1$, for $s=1$
the series for the potential is only asymptotic.   Some explicit results are
summarized in the Appendix for this special $s=1$ case.

\section{Formalism}\label{section2}

Since the functional approach is unconventional and not part of the standard
syllabus for classical mechanics, to make the discussion self-contained and
accessible to the reader, it is necessary to \emph{briefly} review some
earlier work on this subject, especially that previously presented in
\cite{cv, cz1,cz2,cz3}. Of course, the number of known results in the theory
of functional methods is considerable, and the mathematical literature on the
subject is vast.  For a thorough survey of those results and an excellent
guide to that literature, see~\cite{k}.

\subsection{The problem}

Given a discrete map (unit time step, discrete scale change, etc.),%
\begin{gather*}
x\mapsto f_{1}(  x)   ,
\end{gather*}
how do we interpolate (for intermediate times, scales, etc.)?  Such an
interpolation is needed to interpret this dynamical system as a continuous
f\/low (Hamiltonian, renormalization, etc.) sampled discretely.

As we have emphasized in recent papers \cite{cz1,cz2,cv,cz3}, there is an
elegant method, based on the work of Ernst Schr\"{o}der~\cite{s}, that
exploits analyticity in $x$ to f\/ind an interpolation in~$t$.

\subsection{The method}

Schr\"{o}der's functional equation is
\begin{gather*}
s\Psi(  x,s)  =\Psi(  f_{1}(  x,s)  ,s)   ,
\end{gather*}
for some parameter (eigenvalue) $s$, with $\Psi$ to be determined.  So
$f_{1}$ is the \emph{functional conjugation} of the eigen-scale-parameter~$s$,
\begin{gather*}
f_{1} (  x,s)  =\Psi^{-1}(  s\Psi(  x,s)
,s )    ,
\end{gather*}
where the inverse function $\Psi^{-1}$ obeys Poincar\'{e}'s equation
\cite{po},%
\begin{gather*}
\Psi^{-1} (  sx,s )  =f_{1} \big(  \Psi^{-1} (  x,s )
,s \big)    . 
\end{gather*}
The $n$th iterate of the functional equation gives
\begin{gather*}
s^{n}\Psi (  x,s )  =\Psi (  f_{1} (  \dots f_{1} (
f_{1} (  x,s )  ,s )  \dots,s )  ,s )    ,
\end{gather*}
with $f_{1}$ acting $n$ times, and thus the $n$th order functional composition
is also given by a functional conjugation,
\begin{gather*}
f_{n} (  x,s )  \equiv f_{1} (  \dots f_{1} (  f_{1} (
x,s )  ,s )  \dots,s )  =\Psi^{-1} \big(  s^{n} \Psi (
x,s )  ,s \big)   .
\end{gather*}
In principle, with $\Psi$ in hand for a particular problem, any functional
composition of $f_{1}$ can be constructed.

\subsection{The solution}

A continuous interpolation between the integer points is then given by
functional conjugation, for \emph{any}~$t$,
\begin{gather*}
f_{t} (  x,s )  =\Psi^{-1}\big(  s^{t} \Psi (  x,s )
,s\big)  .
\end{gather*}
This can be a well-behaved and single-valued function of~$x$ and~$t$ provided
that $\Psi^{-1} (  x,s )  $ is a well-behaved, single-valued function
of $x$,  even though $\Psi (  x,s )  $ might be, and typically is,
multi-valued.

For generic \emph{real} values of $\Psi$ appearing in the argument of
$\Psi^{-1}$ then:
\begin{itemize}\itemsep=0pt
\item When $s$ is real and $s>1$, the $t\rightarrow\infty$ behavior for
$f_{t} (  x,s )  $ is determined by the large argument asymptotic
behavior of $\Psi^{-1}$.

\item When $s$ is real and $0<s<1$, the $t\rightarrow\infty$ behavior for
$f_{t} (  x,s )  $ is determined by the behavior of $\Psi^{-1}$ near
the origin.

\item When $s$ has a phase, $f_{t} (  x,s )  $ will be complex for
generic $t$.
\end{itemize}

 However, regarding this last point, although it may well have been
of some interest to parti\-ci\-pants at the workshop on \emph{Supersymmetric
Quantum Mechanics and Spectral Design}, here I~will consider only real $s>0$.

\subsection{An interpretation}

The interpolation can be envisioned as the trajectory of a particle, evolving
from initial $x\equiv  x (  t )  \big\vert _{t=0}$.  Thus
\begin{gather*}
x (  t )  =f_{t} (  x,s )   ,
\end{gather*}
where the particle is moving under the inf\/luence of a potential according to
Hamiltonian dynamics. The velocity of the particle is then found by
dif\/ferentiating with respect to~$t$,
\begin{gather*}
\frac{dx (  t )  }{dt}= (  \ln s )   s^{t} \Psi (
x )   (  \Psi^{-1} (  s^{t}\Psi (  x )   )
 )  ^{\prime}  ,
\end{gather*}
where any dependence of $\Psi$ on $s$ is implicitly understood.   Therefore,
the velocity will inherit and exhibit any multi-valued-ness possessed by
$\Psi (  x )  $.

\subsection{The potential}

Up to an additive constant, in suitable mass units, $V(  x)  $ may
be def\/ined by~\cite{cz1,cz2},
\begin{gather*}
V (  x )  \equiv-v^{2} (  x )  ,
\end{gather*}
and its $x$ dependence follows from that of the velocity prof\/ile of the
interpolation,%
\begin{gather}
v(  x)  \equiv\left.  \frac{df_{t}( x)}
{dt}\right\vert _{t=0}=\frac{\ln s}{\frac{d}{dx}\ln\Psi(  x)  } .
\label{v}
\end{gather}
The potential $V\equiv-v^{2}$ then \emph{emerges} as a quadratic in $\Psi
/\Psi^{\prime}$,%
\begin{gather}
V(  x)  =-(  \ln s)  ^{2} \left(  \frac{\Psi (
x )  }{\Psi^{\prime} (  x )  }\right)  ^{2} , \label{V}
\end{gather}
where any dependence of $V$ or $\Psi$ on $s$ is again implicit.

\subsection[A functional equation for $V$]{A functional equation for $\boldsymbol{V}$}

The potential can also be determined directly from the functional equation~\cite{cv} it inherits from~$\Psi$. That is,
\begin{gather}
V (  f_{1} (x,s )  ,s )  =\left(  \frac{d}{dx}f_{1}(
x,s)  \right)  ^{2}V(  x,s)  . \label{FcnVEqn}
\end{gather}
If the discrete map possesses a f\/ixed point, we may attempt to solve this
functional equation for~$V$ by series in~$x$ about that f\/ixed point.

In general, if the series can be constructed, it will of course have a f\/inite
radius of convergence. However, often the series result can then be
continued to other $x$ by making use of the functional equation (a technique
very familiar, e.g., for the~$\Gamma$ and $\zeta$ functions) so that accurate
values for the potential can often be obtained for all~$x$, for any~$s$.
 Such series and continuation techniques may also be used in many, if not all
cases, to determine~$\Psi$ and~$\Psi^{-1}$ to arbitrary precision, in principle.

\subsection{An aside}

For the corresponding supersymmetric systems,
\begin{gather*}
V_{\pm} (  x )  \equiv-v^{2} (  x )  \pm iv^{\prime} (
x )  .
\end{gather*}
This is a complex potential, even for real $s>0$. Despite the emphasis
placed on such systems at the workshop on \emph{Supersymmetric Quantum
Mechanics and Spectral Design}, Centro de Ciencias de Benasque Pedro Pascual,
where this paper was presented, I will not discuss this any further here.

\section{Elementary examples}\label{section3}

\subsection[The Beverton-Holt model]{The Beverton--Holt model}

\begin{figure}[t]\centering
\includegraphics[width=70mm]{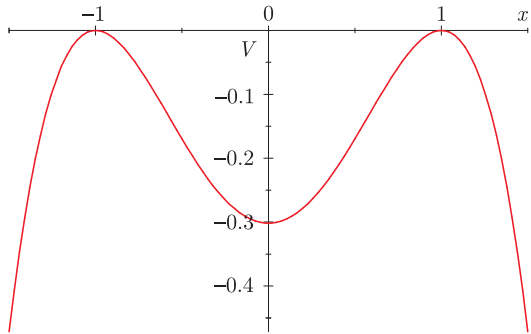}

\caption{An inverted Goldstone potential.}\label{Fig1}
\end{figure}

Consider the rational map,%
\begin{gather}
x\mapsto f_{1} (  x )  =\frac{2x+1}{x+2} , \label{BHMap}
\end{gather}
for $-1\leq x\leq1$. This is known as the classic Beverton--Holt (BH)
discrete-time population model \cite{b}, to some\footnote{The model was also
discussed in an earlier paper on population dynamics: \S~4.3 of~\cite{sk}~--
the necessary change of variables being most readily apparent upon comparing
equation~(42) in~\cite{sk} with~(\ref{BHTrajectory}) to follow.}, in which context
the population is proportional to $1+x$. But physicists~-- especially
particle physicists~-- know it by another name based on the underlying
potential. To see what this is, consider
\begin{gather}
s \Psi (  x )  =\Psi\left(  \frac{2x+1}{x+2}\right) , \qquad \text{with} \quad s=1/3 . \label{BHFE}
\end{gather}
This functional equation is easily \emph{solved by} \emph{series} about either
f\/ixed point, $x_{\ast}=\pm1$, to give series that are quickly
summed\footnote{Schr\"{o}der's equation, in particular~(\ref{BHFE}), is
homogeneous in $\Psi$. So one is free to multiply $\Psi$ by a constant.
 This has no ef\/fect on $v( x)  $ and $V(  x)$, as
evident in~(\ref{v}) and~(\ref{V}), nor does it change the trajectory, so the
underlying physics is the same.}. Choosing $x_{\ast}=+1$ results in
\begin{gather}
\Psi (  x )      =\frac{2x-2}{x+1} ,\qquad \Psi^{-1} (  x )
=\frac{2+x}{2-x}  ,\label{BHPsi}\\
x (  t )      =\Psi^{-1} \big(  s^{t} \Psi (  x,s )
,s\big)  =\frac{ (  x+1 )  + (  x-1 )  (  1/3 )
^{t}}{ (  x+1 )  - (  x-1 )   (  1/3 )  ^{t}},\label{BHTrajectory}\\
V (  x )      =-v^{2} (  x )  =-\tfrac{1}{4}  (
\ln3 )  ^{2} \big(  1-x^{2}\big)  ^{2}. \label{BHV}
\end{gather}
We immediately recognize $V (  x )  $ as the Goldstone potential
\cite{g,n}\footnote{The trajectory (\ref{BHTrajectory}) is the familiar
``S-curve'' of logistic growth, 
\href{http://en.wikipedia.org/wiki/Logistic_function}{wikipedia.org/wiki/Logistic\_function}.}, albeit inverted as
encountered in instanton problems~\cite{p}, as shown in Fig.~\ref{Fig1}.

While $\Psi$ and $\Psi^{-1}$ as given above have poles, the underlying
potential in this case is an entire function, so it is a trivial step to
consider particle motion not just on the original interval, but on the
complete real line, where $V$ is \emph{unbounded below}. (This is one reason
for my chosen title. Another will arise in the discussion of the logistic
map in the next section.)

To be thorough, we note that Schr\"{o}der's method of analysis also works for
an inverted quartic potential with only \emph{one} real extremum. However,
in this case the relevant discrete map and continuous f\/low would also have
purely \emph{imaginary} f\/ixed points. To understand this, consider $v(x)  =1+x^{2}$ with corresponding potential
\begin{gather}
V(x)  =-\big(  1+x^{2}\big)  ^{2}  . \label{Quartic}
\end{gather}
Construction of the zero-energy trajectory, and a $\Psi$ function and it's
inverse, follow immediately for the system governed by (\ref{Quartic}). Thus%
\begin{gather}
\Psi (  x )      =\exp (  \arctan x )  ,\qquad \Psi
^{-1} (  x )  =\tan (  \ln x )  ,\label{QuarticPsi}\\
x (  t )      =\Psi^{-1}\big(  e^{t} \Psi (  x )  \big)
=\tan (  t+\arctan x )  . \label{QuarticTrajectory}
\end{gather}
Similarities to the BH case are highlighted by the discrete time-step map, now
given by%
\begin{gather}
x\mapsto f_{\pi/4} (  x )  =\tan\left(  \frac{\pi}{4}+\arctan
x\right)  =\frac{1+x}{1-x}  , \label{QuarticMap}
\end{gather}
to be compared to~(\ref{BHMap}). The corresponding functional equation is%
\begin{gather}
e^{\pi/4} \Psi (  x )  =\Psi\left(  \frac{1+x}{1-x}\right)  ,
\label{QuarticFE}%
\end{gather}
and is easily checked to be satisf\/ied by $\Psi$ in (\ref{QuarticPsi}) by use
of the addition formula for the tangent.

It is interesting\footnote{I thank the anonymous referee for raising this
issue.} that the discrete map in~(\ref{QuarticMap}) has period four:
 $f_{\pi/4}\circ f_{\pi/4}\circ f_{\pi/4}\circ f_{\pi/4}=\operatorname*{id}$,
i.e.\ \emph{the identity map}, so that $f_{\pi/4} (  f_{\pi
/4} (  f_{\pi/4} (  f_{\pi/4} (  x )   )   ))  =x$. Naively, this would seem to require that the eigenvalue in
Schr\"{o}der's equation must be a fourth root of unity, thus either $\pm1$ or~$\pm i$, and not $e^{\pi/4}$ as given in~(\ref{QuarticFE}). This warrants some explanation.

The solution $\Psi (  x )  =\exp (  \arctan x )  $ was
obtained by solving Schr\"{o}der's equation for continuous time, namely,
\begin{gather}
e^{t}\Psi (  x )  =\Psi (  x (  t )   )  ,
\label{ContinuousQuarticFE}
\end{gather}
where $x\equiv   x (  t )   \vert _{t=0}$ and where the
time-dependent particle trajectory to be used in~(\ref{ContinuousQuarticFE})
is the elementary, right-moving, zero-energy solution~(\ref{QuarticTrajectory}) for the inverted quartic potential~(\ref{Quartic}). However, this
potential deepens so rapidly that a particle moving according to~(\ref{QuarticTrajectory}) always reaches $x=+\infty$ in a \emph{finite} time
starting from any initial $x>0$, namely, $t_{x\Rightarrow\infty}=\frac{\pi}%
{2}-\arctan x$. Of course, this time depends on the initial position but it
always satisf\/ies $t_{x\Rightarrow\infty}<$ $\frac{\pi}{2}$ for any initial
position satisfying $0<x<\infty$.

Now, at this juncture you could adopt the narrow point of view that it makes
\emph{no} physical sense to consider any solution for larger times
$t>t_{x\Rightarrow\infty}$. This would preclude taking $t=\pi$ as required
to produce $f_{\pi/4}\circ f_{\pi/4}\circ f_{\pi/4}\circ f_{\pi/4}=f_{\pi
}=\operatorname*{id}$. And that would be the end of our story.

On the other hand, if you were just a bit more f\/lexible in your thinking, you
might consider the particle to be moving on a circle, albeit of
\emph{infinite} radius, and \emph{identify} the point $x=+\infty$ with the
point $x=-\infty$. Then for time $t>t_{x\Rightarrow\infty}$ the particle's
zero-energy motion would continue, except that the particle would be coming in
from $x=-\infty$ and approaching $x=0$ from the left. But, for zero total
energy in the inverted quartic potential (\ref{Quartic}), it would again take
the particle only a f\/inite time to go from $x=-\infty$ to $x=0$, namely,
$t_{-\infty\Rightarrow0}=\frac{\pi}{2}$, and then the particle would continue
moving to the right so that it would reach its original, initial position in
an additional time $t_{0\Rightarrow x}=\arctan x$.

In this second point of view, for the particle to carry out one
circumnavigation of the inf\/inite circle and return to its original position,
the required total time $t_{\circlearrowright}$ is \emph{finite}, due to the
presence of the negative, unbounded potential. That is to say,
\begin{gather*}
t_{\circlearrowright}=t_{x\Rightarrow\infty}+t_{-\infty\Rightarrow
0}+t_{0\Rightarrow x}=\pi.
\end{gather*}
This is independent of the initial $x$, and it corresponds precisely to the
statement that $f_{\pi/4}\circ f_{\pi/4}\circ f_{\pi/4}\circ f_{\pi/4}\left(
x\right)  =f_{\pi}\left(  x\right)  =x$, the identity map.

Moreover, since the particle's instantaneous transition from $x=+\infty$ to
$x=-\infty$ is, mathe\-matically, just a change from one branch of the $\arctan$
function to the next, for the solution under consideration, Schr\"{o}der's
equation is indeed satisf\/ied when $t=\pi$ but only in the following sense:
\begin{gather*}
e^{\pi}\Psi_{n\text{th branch}} (x)  =\Psi_{n+1\text{st branch}} (x)  , 
\end{gather*}
where $\Psi_{n\text{th branch}} (  x )  =\exp (  \arctan
_{n\text{th branch}} (  x )   )  =\exp (  n\pi
+\arctan_{\text{principal branch}} (  x )   )  $.

Perhaps a better way to describe all this is to project a circle of unit
diameter onto a tangent line. Then a zero-energy trajectory along the
$x$-axis def\/ined by that tangent, in an inverted quartic potential as
described above, corresponds precisely to \emph{constant} angular velocity
($\omega=1$) around the circle. This is the entire content of~(\ref{QuarticTrajectory}). In that expression the argument $\theta=t+\arctan
x$ is just the angle between the circle's diameter perpendicular to the
tangent and a~line from the point on the circle farthest from the tangent to
the point at position $x$ along the tangent line. When re-expressed as a
function of that angle, the particle's energy functional becomes
\begin{gather*}
E[  x]  =(  dx/dt)  ^{2}-\big(  1+x^{2}\big)
^{2}\ \Longrightarrow\ E[  x(  \theta)  ]  =\big(
(  d\theta/dt)  ^{2}-1\big)  /\cos^{4}\theta.
\end{gather*}

To conclude this subsection, it is appropriate to point out that both the
models described here are PT symmetric, i.e.\ invariant under $x\mapsto-x$ and
$t\mapsto-t$. And indeed, in keeping with one of the prevalent themes of the
workshop on \emph{Supersymmetric Quantum Mechanics and Spectral Design}, the
quartic model's potential (\ref{Quartic}) dif\/fers from the BH model's
(\ref{BHV}) simply by a change of time scale and the complexif\/ication:
$1-x^{2}\rightarrow1-(  ix)  ^{2}$.

\subsection{The Skellam model}

The discrete time-stepped map in this case is
\begin{gather}
x\mapsto k\big(  1-e^{-x}\big)  ,\label{SkellamMap}
\end{gather}
where $k$ is a real parameter. For population dynamics, only $0\leq
x<\infty$ is of interest, but the model is well-def\/ined for all the reals (and
$\mathbb{C}$ as well). Fixed points are $x=0$ and $x_{\ast}(  k)
=k+\operatorname{LambertW}(  -ke^{-k})  $. The nontrivial f\/ixed
point $x_{\ast}$ is real and positive for $k>1$, but real and negative for
$0<k<1$, when the appropriate branch of the Lambert function is used. For
$k=1$ there is only the trivial f\/ixed point. This is evident from graphs of
the map, for various $k$.

\begin{figure}[h!]\centering
\includegraphics[width=100mm]{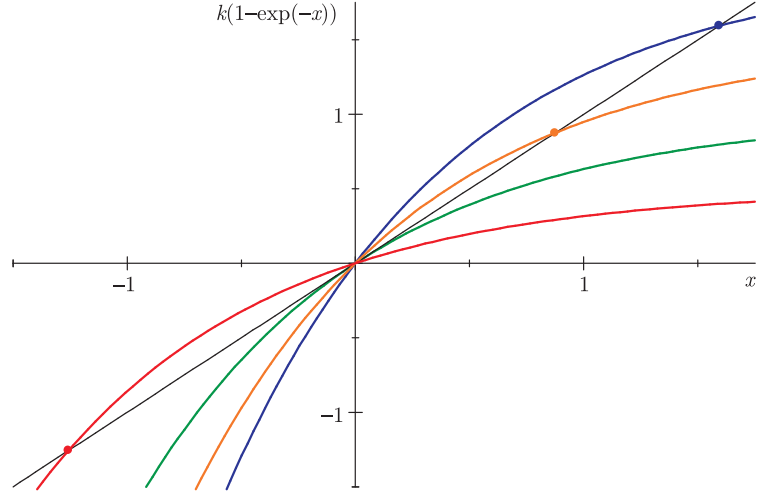}

\caption{Skellam map plotted for various $k$ (namely, $1/2$ in red, $1$ in green, $3/2$
in orange, and $2$ in blue). Dots are the f\/ixed points, $x_{\ast}(k)$.}\label{Fig2}
\end{figure}

This model of population dynamics f\/irst appeared in equation~(39), \S~4 of~\cite{sk}. (After rescaling variables: Skellam's $\Gamma H\chi=x$ here,
and his $\Gamma H=k$. In fact, in his subsequent discussion of \S~4.3,
Skellam anticipated the Beverton--Holt model by approximating the exponential
in (\ref{SkellamMap}) for small $x$: $\exp (  -x )  \approx (
1-x/2 )  / (  1+x/2 )  $, the dif\/ference being $\frac{1}%
{12}x^{3}-\frac{1}{12}x^{4}+\frac{13}{240}x^{5}+O (  x^{6} )$.)
 The various functional equations associated with (\ref{SkellamMap}) are:
\begin{gather}
k \Psi (  x,k )      =\Psi\big(  k\big(  1-e^{-x}\big)
,k\big)  ,\label{SkellamSchroeder}\\
\Psi^{-1} (  kx,k )      =k  \big(  1-e^{-\Psi^{-1} (
x,k )  } \big)  ,\label{SkellamPoincare}\\
k^{2}e^{-2x} V (  x,k )      =V \big(  k \big(  1-e^{-x} \big)
,k\big)  .\label{SkellamPotential}
\end{gather}
It is straightforward to obtain the f\/irst few terms in a series solution of
these functional equations. Unfortunately, unlike the BH case, it is not
possible to express the series in closed-form as known functions, so far as I
am aware. Nevertheless, the series can be used to compute the potentials numerically.

For example, the potential~$V$, expanded about~$0$, is:
\begin{gather}
V (  x,k )  =- (  \ln k )  ^{2}\bigg(  x^{2}-\frac{1}%
{k-1}x^{3}-\frac{1}{12}\frac{k-7}{ (  k+1 )   (  k-1 )
^{2}}x^{4}\nonumber\\
\phantom{V (  x,k )  =}{} +\frac{1}{4\big(  k+k^{2}+1\big)   (  k-1 )  ^{2}}
x^{5}+O\big(  x^{6}\big)  \bigg)  . \label{Vseries}
\end{gather}
An alternate series solution for the potential, expanded about $x_{\ast}$, is:
\begin{gather}
V (  x,k )  =- (  \ln\ell )  ^{2}\bigg(  y^{2}-\frac{1}
{\ell-1}y^{3}-\frac{1}{12}\frac{\ell-7}{ (  \ell+1 )   (
\ell-1 )  ^{2}}y^{4}\nonumber\\
\phantom{V (  x,k )  =}{} +\frac{1}{4\big(  \ell+\ell^{2}+1\big)   (
\ell-1 )  ^{2}}y^{5}+O\big(  y^{6}\big)  \bigg)  ,
\label{Vseriesalternate}%
\end{gather}
where $y\equiv x-x_{\ast}$ and $\ell\equiv k-x_{\ast}=-\operatorname{LambertW}\left(  -ke^{-k}\right)  $. And yes, the coef\/f\/icients in these two series
\emph{do} have exactly the same form.

While the functional equation for $V$ does not determine the normalization,
the relative normalization for the two series~(\ref{Vseries}) and
(\ref{Vseriesalternate}) has been f\/ixed by the requirement that they match on
the interval $0<x<x_{\ast}$. With the relative normalization as given, there
is also a \emph{duality} between the right-hand sides of~(\ref{Vseries}) and~(\ref{Vseriesalternate}) to the orders shown: $x,k\Longleftrightarrow
y,\ell$. Under this interchange the expressions are \emph{identical in
form}, as already noted. This equivalence between the two series expansions
can be established \emph{to all orders} by inspection of the corresponding
recursion relations for the series coef\/f\/icients.

\looseness=-1
For all positive $k>1$ the potentials for the Skellam model are qualitatively
similar to the inverted Goldstone potential, just not as symmetrical. Outside
the regions between the two real f\/ixed points, which are local maxima, the
potentials are apparently unbounded below. Between the f\/ixed points the
potentials have a single local minimum at $x_{\min}\left(  k\right)  \approx
x_{\ast}/2$. For any $k>1$, the leading $x^{2}$ and $y^{2}$ behavior of~$V$
in~(\ref{Vseries}) and~(\ref{Vseriesalternate}) show that an inf\/inite amount
of time is required for a zero-energy trajectory to reach the lower and upper
f\/ixed points, respectively, starting from somewhere else. Consequently, the
Schr\"{o}der function constructed about $x=0$, with $\Psi (  x )
=x+O \left(  x^{2} \right)  $, will truly diverge to $+\infty$ as $x\rightarrow
x_{\ast}$. Another way to say this is that $\Psi^{-1}$ will asymptote to a
constant as its argument becomes large, $\lim\limits_{z\rightarrow\infty}\Psi
^{-1} (  z,k )  =x_{\ast}$, when $k>1$.

For positive $k<1$, the Skellam model potentials also have structure
qualitatively similar to the inverted Goldstone potential, except that the
nontrivial local maximum and the local minimum are at negative $x$. For
$k=1$ the potential is best def\/ined as a limit, $V (  x,1 )
=\lim\limits_{k\rightarrow1}V (  x,k )  = -\frac{1}{4}x^{4}%
-\frac{1}{12}x^{5}+O \left(  x^{6} \right)  $. However, in this latter case
the series for $V (  x,1 )  $ is probably not convergent, but
asymptotic instead. (We discuss in some detail a similar situation for a
special case of the logistic map, in the Appendix. The corresponding
analysis of the $k=1$ Skellam model is left as an exercise for the reader.)

We plot the potentials in the region between the two f\/ixed points for a few
representative values of $k>1$, using the series expansions~(\ref{Vseries})
and~(\ref{Vseriesalternate}), only extended to tenth order in~$x$ and~$y$.
 This order is suf\/f\/icient for any errors in the graphs to be insignif\/icant,
and indiscernible.
\begin{figure}[h!]\centering
\includegraphics[width=105mm]{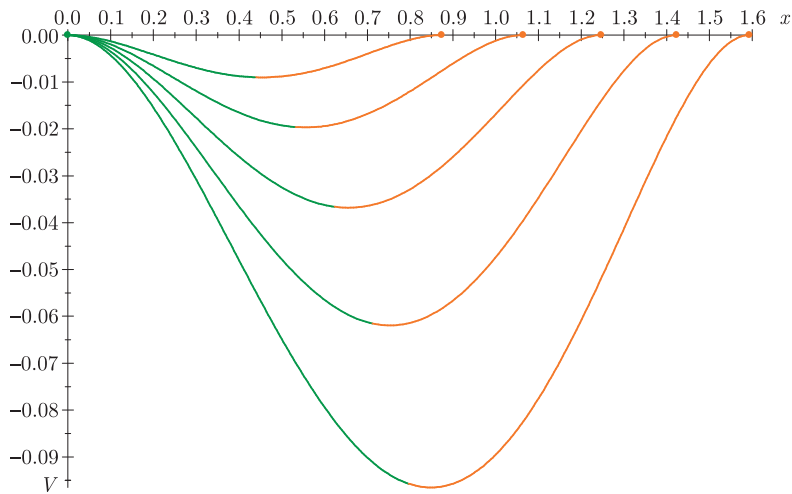}

\caption{Skellam model potentials, $V (  x,k )  $, for
$k=3/2,\, 13/8,\, 7/4,\, 15/8$, and $2$ (as successively lower curves) from the
series about $0$ (in green) joined at $x_{\ast}/2$ to the series about
$x_{\ast}$ (in orange). Dots on the $x$ axis are the f\/ixed points.}
\end{figure}

The continuous trajectories for the Skellam model can also be computed
numerically, although they are not known in simple, closed form. The most
direct way to do this is to compute the inverse of the Schr\"{o}der function
from~(\ref{SkellamPoincare}), say as series, and then re-use that functional
equation to improve upon the numerical results obtained from the series.
 Since this article is mainly about the potentials that underlie the
trajectories, I~will not present here numerical results for~$\Psi^{-1}$. But
I~will discuss the series solution of~(\ref{SkellamPoincare}) for generic~$k$.

The functional equation for $\Psi^{-1}$ is interesting in its own right. Let
$\Psi^{-1}\left(  x,k\right)  =-f\left(  -x,k\right)  $, then
(\ref{SkellamPoincare}) becomes%
\begin{gather}
f (  kx,k )  =k\big(  e^{f (  x,k )  }-1\big)  .
\label{BellSkellam}%
\end{gather}
Consider a formal series solution:%
\begin{gather}
f (  x,k )  =\sum_{j=1}^{\infty}\frac{b_{j}x^{j}}{j!}  .
\label{SkellamSeries}
\end{gather}
The exponential
formula\footnote{\url{http://en.wikipedia.org/wiki/Exponential_formula}.},
\begin{gather*}
\exp\left(  \sum_{j=1}^{\infty}\frac{b_{j}x^{j}}{j!}\right)  =\sum
_{n=0}^{\infty}\frac{B_{n} (  b_{1},\dots,b_{n} )  }{n!}x^{n}  ,
\end{gather*}
where $B_{n} (  b_{1},\dots,b_{n} )  $ are the complete
multi-variate Bell
polynomials\footnote{\url{http://en.wikipedia.org/wiki/Bell_polynomial\#Complete_Bell_polynomials}.},
permits us to write the functional equation (\ref{BellSkellam}) as
\begin{gather*}
\sum_{j=1}^{\infty}\frac{b_{j}k^{j}x^{j}}{j!}=k\sum_{n=1}^{\infty}\frac
{B_{n} (  b_{1},\dots,b_{n} )  }{n!}x^{n}  ,
\end{gather*}
where it is understood that the $b_{j}$ are $k$-dependent. Thus the
coef\/f\/icients of $x^{n}$ give the recursion relation,
\begin{gather}
k^{n-1}b_{n}=B_{n} (  b_{1},\dots,b_{n} )  . \label{Precursion}
\end{gather}
Since all of these complete Bell polynomials depend on their last argument
only as $B_{n} (  b_{1},\dots,b_{n} )$  $=b_{n}+\cdots$, with no other
dependence on $b_{n}$, we can rewrite (\ref{Precursion}) in a technically
sweeter form:
\begin{gather}
b_{n}=\frac{1}{k^{n-1}-1} B_{n} (  b_{1},b_{2},\dots,b_{n-1},0 )
  . \label{SkellamRecursion}
\end{gather}
Moreover, since the $B_{n}$ are homogeneous of degree $n$ under the
multi-variate rescaling $b_{k}\rightarrow r^{k}b_{k}$, without loss of
generality we may choose $b_{1}=1$, thereby setting the scale of $x$ in the
expansion (\ref{SkellamSeries})\footnote{From (\ref{SkellamRecursion}) one
can show the series~(\ref{SkellamSeries}) converges for all $x$ when $k>1$.}.

Because of the prefactor in (\ref{SkellamRecursion}), we see that each
$b_{n+1}$ is a polynomial in $k$ multiplied by a factor $1/ \prod\limits_{m=1}^{n}
\left(  k^{m}-1\right)  $, and thus singular as $k\rightarrow1$. More
precisely, I f\/ind $b_{n+1} (  k )  \underset{k\rightarrow1}{\sim
}\frac{1}{ (  k-1 )  ^{n}} (  n+1 )  !\left(  \frac{1}%
{2}\right)  ^{n}$. So, writing%
\begin{gather*}
b_{n+1} (  k )  =\frac{p_{n} (  k )  }{\prod\limits_{m=1}^{n}
\left(  k^{m}-1\right)  }  ,
\end{gather*}
the resulting $p_{n} (  k )  $ are polynomials in $k$ of order
$n (  n-1 )  /2$. With the choice of initial condition $b_{1}=1$,
these polynomials have only positive integer coef\/f\/icients. The f\/irst few
are:
\begin{gather*}
p_{1} (  k )      =1  ,\qquad
p_{2} (  k )      =2+k  , \qquad
p_{3} (  k )      =6+6k+5k^{2}+k^{3}  , \\
p_{4} (  k )      =24+36k+46k^{2}+40k^{3}+24k^{4}+9k^{5}%
+k^{6}  , \\
p_{5} (  k )      =120+240k+390k^{2}\!+480k^{3}\!+514k^{4}\!+416k^{5}\!
+301k^{6}\!+160k^{7}\!
+64k^{8}\!+14k^{9}\!+k^{10}  , \\
p_{6} (  k )      =720+1800k+3480k^{2}+5250k^{3}+7028k^{4}%
+8056k^{5}+8252k^{6}+7426k^{7} \\
\phantom{p_{6} (  k )      =}{}   +5979k^{8}+4208k^{9}+2542k^{10}+1295k^{11}+504k^{12}+139k^{13}%
+20k^{14}+k^{15}  ,
\end{gather*}
etc. From these, I infer two particular values of the polynomials, for any
$n$,
\begin{gather*}
   p_{n} (  k )   \vert _{k=0}=n!  ,\qquad
p_{n} (  k )   \vert _{k=1}=n! (  n+1 )  ! \left(
\frac{1}{2}\right)  ^{n}  . 
\end{gather*}
In fact, the coef\/f\/icients of the lowest two and highest two powers of~$k$ are
easily determined for any $n$,
\begin{gather*}
p_{n} (  k )  =n!+\frac{1}{2} (  n-1 )  n!k+\cdots+\frac
{1}{2} (  n+2 )   (  n-1 )  k^{n (  n-1 )
/2-1}+k^{n (  n-1 )  /2}  .
\end{gather*}
This result may be established by induction, and probably extended to other
coef\/f\/icients as well, given enough time and interest.

In terms of the $p_{n}$s, the formal series solution for $\Psi^{-1}$ is%
\begin{gather*}
\Psi^{-1}(  x,k)  =\sum_{n=1}^{\infty}\frac{(  -1)
^{n-1}b_{n}(  k)  x^{n}}{n!}=x+x\sum_{n=1}^{\infty}\frac{1}{(n+1)  !}\frac{p_{n}(  k)  x^{n}}{%
\prod\limits_{i=1}^{n}
\left(  1-k^{i}\right)} . 
\end{gather*}
We may also express this in terms of deformed, $k$-bracket
factorials\footnote{\url{http://en.wikipedia.org/wiki/Q-bracket\#Relationship_to_the_q-bracket_and_the_q-binomial}.},
\begin{gather*}
[  n]  _{k}!=
\prod\limits_{m=1}^{n}
[  m]  _{k} ,\qquad   [  m ]  _{k}=\frac{k^{m}-1}{k-1} .
\end{gather*}
This makes it clear that the series is actually in terms of the rescaled
variable $x/ (  1-k )  $. Thus,%
\begin{gather*}
\Psi^{-1} (  x,k )  =x+x\sum_{n=1}^{\infty}\frac{1}{ (
n+1 )  !}\frac{p_{n} (  k )   \left(  \frac{x}{1-k}\right)
^{n}}{ [  n ]  _{k}!}  .
\end{gather*}
As a special case, this has a simple limit as $k\rightarrow0$, namely,
\begin{gather*}
\lim_{k\rightarrow0}\big(  \Psi^{-1} (  x,k )  \big)
=x+x\sum_{n=1}^{\infty}\frac{1}{ (  n+1 )  !}\frac{p_{n} (
0 )   x^{n}}{1}
   =x+x\sum_{n=1}^{\infty}\frac{1}{n+1}x^{n}=-\ln (  1-x )  .
\end{gather*}
Another special case yields a direct connection to the BH model, also by
taking a limit. Again interchanging limit with summation we f\/ind{\samepage
\begin{gather*}
\lim_{k\rightarrow1}\left(  \frac{1}{1-k}~\Psi^{-1} (   (  1-k )
x,k )   \right)      =x+x\sum_{n=1}^{\infty}\frac{1}{(  n+1)
!}\frac{p_{n}(  1)  x^{n}}{n!}\nonumber\\
 \hphantom{\lim_{k\rightarrow1}\left(  \frac{1}{1-k}~\Psi^{-1} (   (  1-k )
x,k )   \right)}{}
   =x+x\sum_{n=1}^{\infty}\left(  \frac{x}{2}\right)  ^{n}=\frac{2x}
{2-x}   =\Psi_{\rm BH}^{-1} (  x )  -1 ,
\end{gather*}
where $\Psi_{\rm BH}^{-1}$ is the inverse Schr\"{o}der function in~(\ref{BHPsi}).}

A numerical study of $\Psi^{-1}$ and the corresponding continuous trajectories
for the Skellam model for general~$k$ will have to be presented elsewhere.
 For $k\neq1$, the trajectories are qualitatively the same as the
``S-curves'' of the BH model, as can be
inferred just from the qualitative similarity of the inverted Goldstone
potential to the Skellam model potentials, such as those exhibited in the
Figure above.

Further features of the functional formalism will now be explored for a
dif\/ferent, perhaps better-known class of models.

\section{The logistic map}\label{section4}

A well-studied and interesting map is given by \cite{ce,d}
\begin{gather*}
x\mapsto s x (  1-x )  ,
\end{gather*}
on the unit interval, $x\in [  0,1 ]  $, for parameter values
$s\in [  0,4 ]  $. The maximum of the map is $s/4$,  obtained from
$x=1/2$, so without loss of any essential features, for a given $s$, we need
only consider $x\in [  0,s/4 ]  $. The map has f\/ixed points at
$x=0$ and at $x_{\ast}=1-1/s$.

Iteration of this map shows a wealth of interesting, well-known
features\footnote{\url{http://en.wikipedia.org/wiki/Logistic_map}.}.
\begin{itemize}\itemsep=0pt
\item With $0<s<1$, $x_{n}\underset{n\rightarrow\infty}{\rightarrow}0$
independent of the initial $x$.

\item With $1<s\leq2$, $x_{n}\underset{n\rightarrow\infty}{\rightarrow}%
1-\frac{1}{s}$, monotonically.

\item With $2<s\leq3$, $x_{n}\underset{n\rightarrow\infty}{\rightarrow}%
1-\frac{1}{s}$, oscillatorially.

\item With $3<s\leq1+\sqrt{6}\approx3.45$, $x_{n}$ oscillates between two ($s$
dependent) values.

\item With $3.45\lesssim s\lesssim3.54$, $x_{n}$ oscillates between four ($s $
dependent) values. Etc.~-- the so-called ``period-doubling
cascade''.

\item At $s\approx3.57$ the period-doubling cascade ends, and the $x_{n}$
sequence becomes chaotic. Slight variations in the initial $x$ yield
dramatically dif\/ferent results for the sequence of $x_{n}$.

\item Chaotic behavior holds for almost all larger $s$ up to and including
$s=4$.
\end{itemize}

Schr\"{o}der's equation for the logistic map is
\begin{gather*}
s\Psi (  x,s )  =\Psi (  sx (  1-x )  ,s )  .
\end{gather*}
The inverse function satisf\/ies the corresponding Poincar\'{e} equation,%
\begin{gather*}
\Psi^{-1} (  sx,s )  =s\Psi^{-1} (  x,s )   \big(
1-\Psi^{-1} (  x,s )  \big)  .
\end{gather*}
As originally obtained by Schr\"{o}der, there are three closed-form solutions
known, for $s=-2,$ $2,$ and $4$ ($s=-2$ is related to $s=4$ by functional
conjugation):{\samepage
\begin{alignat*}{3}
& \Psi (  x,2 )      =-\frac{1}{2}\ln (  1-2x )
 ,\qquad && \Psi^{-1} (  x,2 )  =\frac{1}{2}\big(  1-e^{-2x}\big)
 ,& \\
& \Psi\big(  x,4\big)      =\big(  \arcsin\sqrt{x}\big)  ^{2}
  ,\qquad && \Psi^{-1} (  x,4 )  =\big(  \sin\sqrt{x}\big)  ^{2}  .&
\end{alignat*}
Note that $\Psi (  x,4 )  $ is multi-valued, but $\Psi^{-1} (
x,4 )  $ is single-valued, for $0\leq x\leq1$.}

While the solutions for $\Psi$ for other values of $s$ are not known in a
simple \emph{closed} form, they may be constructed as convergent series. The
functional method then reveals that there are well-def\/ined expressions for the
continuous time interpolation of the logistic map for \emph{all} $s$ values.

Moreover,  the potentials needed to produce the explicit, continuously
evolving trajectories of the envisioned particle, for all~$s$, can now be
obtained. These potentials had not been obtained prior to our work, so far
as we can tell from the literature. For $s>3$ these potentials are unbounded
below in an interesting way.

A new feature for the potentials so obtained is that they show the
``switchback ef\/fect'': \emph{If and when the
particle reaches a turning point, the potential changes form}. This requires
a bit of explanation.

\subsection[The $s=4$ case in more detail]{The $\boldsymbol{s=4}$ case in more detail}

The trajectory, velocity, and potential are given in this case by%
\begin{gather*}
x (  t )     =\Psi^{-1}\big(  4^{t} \Psi (  x,4 )
,4\big)  =\big(  \sin\big(  2^{t}\arcsin\sqrt{x}\big)  \big)
^{2}\ ,\\
\frac{dx (  t )  }{dt}     = (  \ln4 )  \sqrt{x (
t )   (  1-x (  t )   )  }\arcsin\sqrt{x (
t )  }  ,\\
V (  x (  t )   )      = (  \ln4 )  ^{2}x (
t )   (  x (  t )  -1 )  \arcsin^{2}\sqrt{x (
t )  }  .
\end{gather*}
The last expression appears to be time-translationally invariant. However,
the velocity function has branch points (i.e.\ turning points) at $x (
t )  =0$ and $x (  t )  =1$, so some care is needed to
\emph{determine which branch of the potential function is actually in effect},
at any given time, since the turning points are encountered at f\/inite times.

Initially, the particle is clearly moving in a potential where all the
functions are principal valued for $x>0$,
\begin{gather}
V_{0} (  x )  = (  \ln4 )  ^{2}x (  x-1 )
\arcsin^{2}\sqrt{x} . \label{PrimaryPotential4}
\end{gather}

\begin{figure}[h!]\centering
\includegraphics[width=100mm]{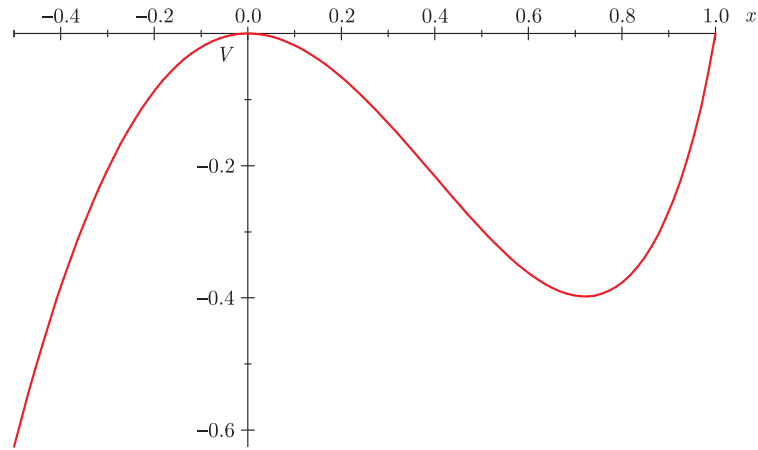}
\caption{The primary potential, $V_{0} (x)$.}\label{Fig4}
\end{figure}

This potential is again unbounded below, as $x\rightarrow-\infty$, \emph{but}
the structure for $x>0$ is more interesting. So, consider the particle to be
moving to the right, with initial $x>0$.

Since the RHS of $V_{0}$ does not have a ref\/lection-free, ``inf\/inity pool'' edge, the zero-energy particle initially
moving to the right will reach $x=1$ in a f\/inite time,
\begin{gather*}
\Delta t_{0} (  x )  =\frac{1}{\ln4}\int_{x}^{1}\frac{dy}
{\sqrt{y (  1-y )  }\big(  \arcsin\sqrt{y}\big)  }=\frac{1}{\ln
2} \ln\left(  \frac{\pi/2}{\arcsin\sqrt{x}}\right)  .
\end{gather*}
It turns around at this time, but its subsequent motion is that of a
zero-energy particle moving to the left in a \emph{modified} potential,
$V_{1} (  x )  $. This follows just from inspection of $v (
t )  $, and setting $V_{1} (  x )  =-v^{2} (  t )  $
after the encounter with the turning point.

This modif\/ied potential does not have ref\/lection-free edges at either side, as
evident in its explicit expression,
\begin{gather*}
V_{1} (  x )  = (  \ln4 )  ^{2}x (  x-1 )   \big(
-\pi+\arcsin\sqrt{x}\big)  ^{2}  .
\end{gather*}

\begin{figure}[h!]\centering
\includegraphics[width=100mm]{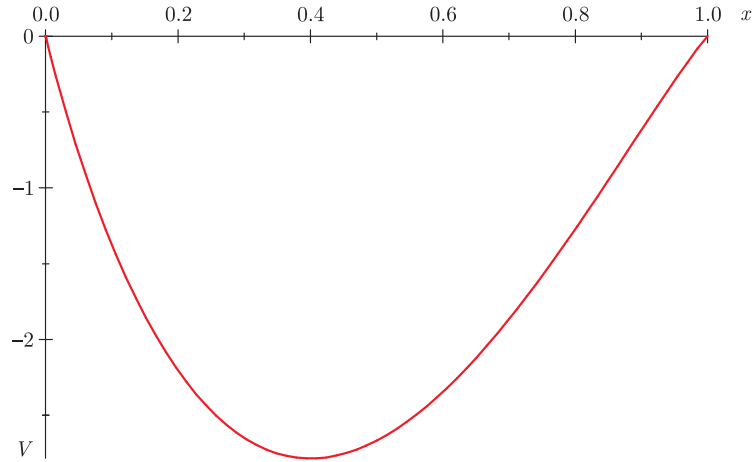}

\caption{The f\/irst switchback potential, $V_{1}(x)$.}\label{Fig5}
\end{figure}

Moving in this modif\/ied negative potential, it now takes the zero-energy
particle a \emph{finite} amount of time to travel from $x=1$ down to $x=0$, as
given by
\begin{gather*}
\Delta t_{1}=1 .
\end{gather*}
Upon reaching $x=0$, the particle is again ref\/lected, and the exact solution
$x (  t )  $ exhibits another change in velocity prof\/ile that keeps
$E=0$, but again alters the potential for the return trip towards $x=1$. The
potential becomes%
\begin{gather*}
V_{2}\left(  x\right)  =\left(  \ln4\right)  ^{2}x\left(  x-1\right)  \left(
\pi+\arcsin\sqrt{x}\right)  ^{2} .
\end{gather*}

\begin{figure}[h!]\centering
\includegraphics[width=100mm]{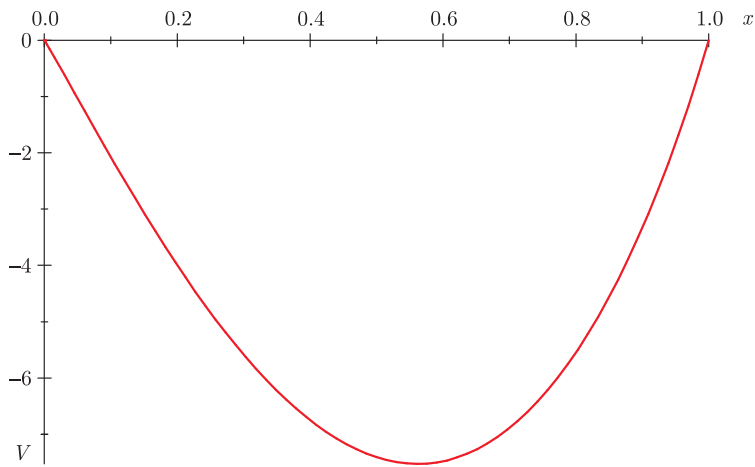}

\caption{The second switchback potential, $V_{2} (  x )  $.}\label{Fig6}
\end{figure}

Notice that the sequence $V_{0}\rightarrow V_{1}\rightarrow V_{2}$ involves
successively deeper, negative potentials.

And so it goes. The motion of the particle between its encounters with the
$P$th and $(  P+1 )$st turning points is determined by the
potential:
\begin{gather*}
V_{P} (  x )  = (  \ln4 )  ^{2}x (  x-1 )  \left(
  ( - 1)  ^{P}\left\lfloor \frac{1+P}{2}\right\rfloor \pi+\arcsin
\sqrt{x}\right)  ^{2}  ,
\end{gather*}
where $\left\lfloor \cdots\right\rfloor $ is the f\/loor function and $\arcsin$
is understood to take principal values. That is to say, the potential
deepens as $P$ increases.

The corresponding velocity prof\/ile speeds up:
\begin{gather*}
v_{P} (  x )  = (  \ln4 )  \sqrt{x (  1-x )
}\left(   (  - 1)  ^{P}\left\lfloor \frac{1+P}{2}\right\rfloor
\pi+\arcsin\sqrt{x}\right)  .
\end{gather*}
The particle will either be traveling to the left, with $v_{P} (
x )  <0$ for odd $P$, or traveling to the right, with $v_{P} (
x)  >0$ for even $P$, with its speed increasing with~$P$. This ef\/fect
is clearly seen upon viewing animations of the $s=4$ trajectories\footnote{\url{http://server.physics.miami.edu/~curtright/Schroeder.html}.}. 

It is instructive to plot $E (  t )  =v^{2} (  x (  t )
 )  +V_{P} (  x (  t )   )  $ versus $t$ for various
$P$, to check $E=0$, as well as to see how the energy would \emph{not} be
conserved if the potentials were \emph{not} switched.

\begin{figure}[h!]\centering
\includegraphics[width=120mm]{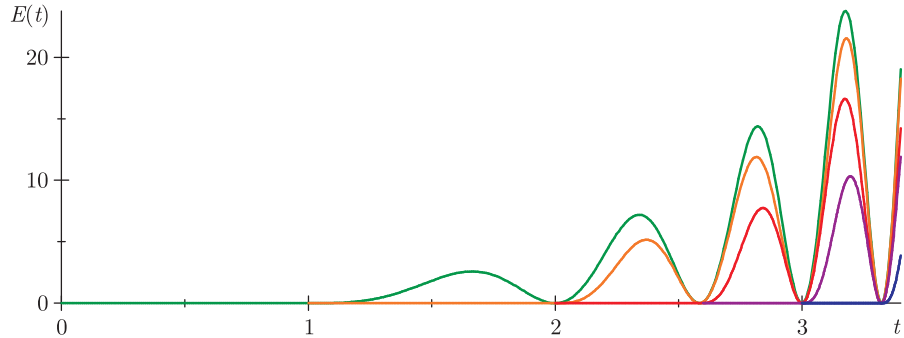}

\caption{$E (  t )  $ for initial $x=1/2$, using potentials $V_{P}$, with
$P=0$, $1$, $2$, $3$, and $4$, as shown in green, orange, red, purple, and
blue, resp.}\label{Fig7}
\end{figure}

For particles with initial $v>0$, another way to picture the zig-zag motion of
the particle is in terms of the total distance traveled. In this point of
view, the successive $V_{P}$ patch together to form a continuous potential
$V(  X)  $ on the real half-line, $X\geq0$. Indeed, the half-line
may be thought of as a \emph{covering manifold} of the unit interval in~$x$,
with the previous multi-valued functions of~$x$ now single-valued functions of
$X$. In this way of looking at the particle's motion, the potential is again
unbounded below (with a downward parabola as envelope), even though at any
time, $x>0$.

\begin{figure}[h!]\centering
\includegraphics[width=100mm]{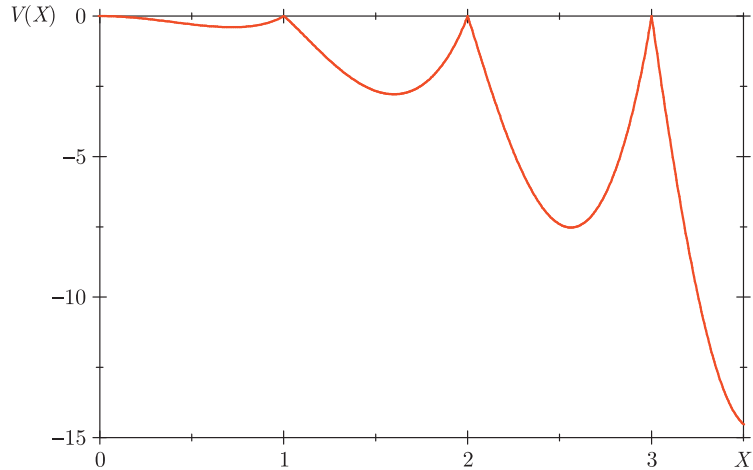}

\caption{$V(X)$ where the total distance traveled by the particle is $X-x (  t )  \vert _{t=0}$.}\label{Fig8}
\end{figure}

Alternatively, the motion may be visualized as a trajectory on the sheets of a
Riemann surface. Consider the particle moving on the complex $x$ plane, and
not just the real line segment $ [  0,1 ]  $, the endpoints of which
are now branchpoints. There are cuts from~$+1$ to~$+\infty$ and from~$0$ to~$-\infty$. The particle f\/irst moves along the real axis, approaches the cut
at $+1$, and then goes around the branchpoint such that
\[
 \sqrt{x}\rightarrow\sqrt{x} , \qquad \sqrt{1-x}\rightarrow-\sqrt{1-x},
\qquad \arcsin\sqrt{x}\rightarrow\pi-\arcsin\sqrt{x}.
\]
The particle returns along the real axis to the origin and encircles
the branchpoint at $0$, such that
\[
 \sqrt{x}\rightarrow-\sqrt{x}, \qquad \sqrt{1-x}\rightarrow\sqrt{1-x},
\qquad \arcsin\sqrt{x}\rightarrow-\arcsin\sqrt{x}.
\]
The particle then goes back to $+1$ and goes around it once more
such that again
\[
\sqrt{x}\rightarrow\sqrt{x}, \qquad \sqrt{1-x}\rightarrow-\sqrt{1-x},
\qquad \arcsin\sqrt{x}\rightarrow\pi-\arcsin\sqrt{x}.
\]
The trajectory continues in this way, f\/lipping signs and adding
$\pi$s according to the formulae for~$V_{P}$ and~$v_{P}$.

\subsection[Series solutions for other values of $s$]{Series solutions for other values of $\boldsymbol{s}$}

The functional equation for the potential that underlies the logistic map is
\begin{gather*}
V (  sx (  1-x )  ,s )  =s^{2} (  1-2x )
^{2}V (  x,s )  .
\end{gather*}
Applying the method of series solution about $x=0$, with initial conditions
$V (  0,s )  =0$, $V^{\prime} (  0,s )   =0$, and
$V^{\prime\prime} (  0,s )  =-2\ln^{2}s$, we f\/ind:
\begin{gather}
V (  x,s )      =- \big(  \ln^{2}s\big)   U (  x,s )
 ,\label{LogisticPotentialSeries}\\
U (  x,s )      =x^{2}\left(  1+\sum_{n=1}^{\infty}a_{n} (
s )   x^{n}\right)  ,\qquad  a_{1}=\frac{2}{1-s}  ,\qquad  a_{2}
=\frac{5-3s}{(  s-1)  ^{2}(  s+1)  } ,\qquad  \dots  ,\nonumber
\end{gather}
provided $s\neq1$. The higher coef\/f\/icients in the expansion are determined
recursively by
\begin{gather*}
a_{n+2}=\frac{1}{\left(  1-s^{n+2}\right)  }\left(  4a_{n+1}-4a_{n}+\sum_{j=1+\left\lfloor \frac{n-1}{2}\right\rfloor }^{n+1} (  -1 )
^{n-j}a_{j}s^{j}\binom{j+2}{n+2-j}\right)  ,
\end{gather*}
for $n\geq1$, where $ \lfloor \cdots \rfloor $ is the f\/loor function.
 This series solves the functional equation for any~$s$, within its radius of
convergence, $R (  s )  $. From numerical studies, that radius
depends on $s$ as follows:
\begin{gather}
R (  s )  =\frac{1}{\lim\limits_{n\rightarrow\infty}\sup\big(
 \vert a_{n} (  s )   \vert ^{1/n}\big)  }=\left\{
\begin{array}{ll}
\dfrac{1}{2} & \text{if}  \ \ 0<s\leq\dfrac{2}{3}  ,\vspace{2mm}\\
\left\vert 1-\dfrac{1}{s}\right\vert  & \text{if}  \ \ \dfrac{2}{3}\leq
s\leq2  ,\vspace{2mm}\\
\dfrac{s}{4} & \text{if} \  \ 2\leq s\leq4 .
\end{array}
\right.  \label{Radii}
\end{gather}
In particular, $R (  4 )  =1$, as is evident from~(\ref{PrimaryPotential4}), while taking~(\ref{Radii}) at face value, $R(1)  =0$, suggesting an asymptotic series. More on the $s=1$ case as a
limit of~(\ref{LogisticPotentialSeries}) can be found in the Appendix.

\subsection{Functional continuation of the series solutions}

The sequence of switchback potentials, i.e.\ the various branches of the
analytic potential function, can be obtained from the functional equation for
the potential. This follows from re-expressing the functional equation as%
\begin{gather}
U_{\pm} (  x )      =s (  s-4x )   U (  x_{\pm} )
  ,\qquad
x_{\pm}     =\frac{1}{2}\pm\frac{1}{2}\sqrt{1-4x/s}  . \label{xPlusMinus}
\end{gather}
One of these potentials ($U_{-}$) reproduces the original series for the
``primary potential'' expanded about $x=0$,
as may be seen by direct comparison of the series or by numerical evaluation.
 Alternatively, the other ($U_{+}$) gives the potential on another sheet of
the function's Riemann surface, hence the f\/irst ``switchback
potential'' for $s>2$. This is easily checked against the
closed-form results for the $s=4$ case.

Since $U_{-}$ has built into it zeroes at both $x=0$ and $x_{\max}=s/4$,
it is actually a more useful form than just the series about $x=0$. When
$2<s\leq4$, the series has radius of convergence $R(  s)  =s/4$.
However, from using $U_{-}$ instead of the direct series, we have
convergence over the whole closed interval, $x\in[  0,s/4]  $,
since then $0\leq\frac{1}{2}-\frac{1}{2}\sqrt{1-4x/s}\leq1/2<R(
s)  $ when $2<s\leq4$. Thus we need only evaluate $U$ appearing in
$U_{-}$ within the region of convergence of its series about zero. That is
to say, f\/irst construct the series for $U$, and then from that series build
$U_{-}$. Finally, identify this with $U_{0}$, the primary potential in the
sequence.

Similarly, $U_{+}$ may be identif\/ied with the f\/irst switchback in the
sequence, $U_{1}$, but for better convergence properties it is useful to
build\ $U_{+}$\ from $U_{0}$ instead of $U$. By doing this when $2<s\leq4$, we have convergence over the whole closed sub-interval, $x\in\left[
\frac{1}{16}s^{2}\left(  4-s\right)  ,s/4\right]  $, with zeroes of $U_{1}$ built-in at the end-points of the interval.

This process may be continued indef\/initely, through successive application of
the basic substitution $U(  x)  \rightarrow s(  s-4x)
U(  x_{\pm})  $. For example, the next set of potentials in the
sequence is $U_{\pm\pm\pm}(  x)  =s(  s-4x)  U_{\pm\pm
}(  x_{\pm})  $, etc. In general, the $n$th iteration of the
procedure gives
\begin{gather}
U_{\underset{\text{$n$ times}}{\underbrace{\pm\pm\cdots\pm}}} (  x )
=s (  s-4x )  U_{\underset{\text{$n-1$ times}}{\underbrace{\pm\pm
\cdots\pm}}} (  x_{\pm} )  . \label{FcnUEqn}
\end{gather}
Finally, at each iteration, we must select appropriate switchback potentials
out of the $2^{n}$ dif\/ferent expressions. In particular, we note that many
of the $U_{\pm\pm\cdots\pm} (  x )  $ will be complex-valued for the
$x$ intervals under consideration, and therefore they are not of immediate
interest since they do not govern the particle's evolution along the real
axis. (The continuation of the particle trajectory into the complex plane is
outside the scope of this paper.)

\subsection{Numerical examples}

Consider the specif\/ic case $s=5/2$. This is representative for $2<s<3$.
 For this case, the turning points converge onto the nontrivial f\/ixed
point $x_{\ast}=1-1/s=3/5$. This is evident in the following graphs. For
$s\leq3$, the potential sequence is actually bounded below, for positive~$x$
(but not for negative $x$). In fact, for $s<3$ the potentials in the
sequence become progressively more shallow for $x>0$ and tend to disappear!

\begin{figure}[h!]\centering
\includegraphics[width=100mm]{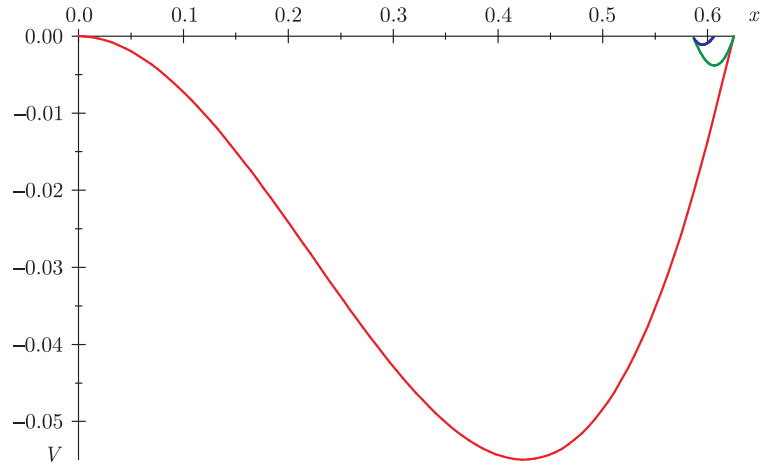}

\caption{Logistic map potentials for $s=5/2$.}\label{Fig9}
\end{figure}

The turning points and the behavior of the potentials near  $x_{\ast}=3/5$ are
shown more clearly below, for the f\/irst three potential switchbacks.

\begin{figure}[h!]\centering
\includegraphics[width=100mm]{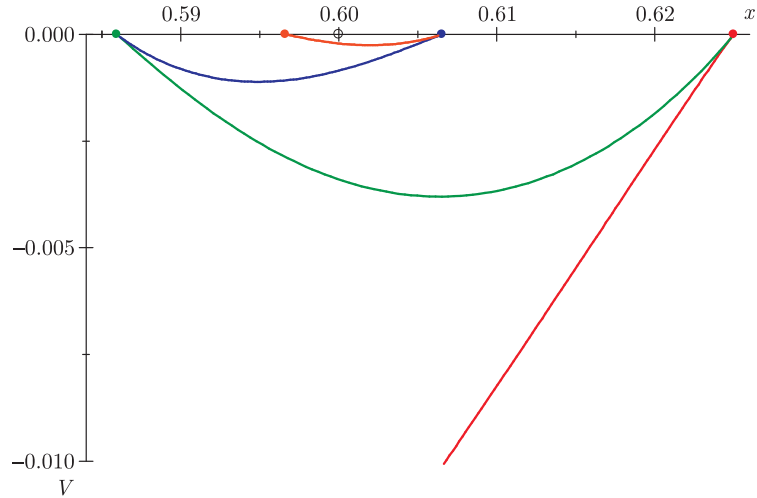}

\caption{Logistic map potentials for $s=5/2$, near the f\/ixed point $x_{\ast}=3/5$.}\label{Fig10}
\end{figure}

As the zero-energy particle moves through this sequence of increasingly
shallow, narrowing potentials, its average speed decreases, giving the
\emph{appearance} of a dissipative system. Nevertheless, even as the
particle motion subsides upon convergence into the f\/ixed point at~$x=3/5$,
energy is rigorously conserved through changes in the potential.

For other values of~$s$, the sequence of potentials can also be accurately
determined using series solutions augmented by functional methods.

Rather than plot the potentials, another useful way to envision the dynamics
is to construct a phase-space trajectory~\cite{cz2}. This is easily done for
those cases where the trajectories are known in closed-form. For cases where
numerical analysis is required, it is also relatively straightforward to
construct graphs of the phase-space curves using the series solutions,
augmented by continuation through use of the functional equations, although
for the phase space the relevant functional equation is that for the momentum
(note that $p=2v$ for the mass units we have chosen). This is the
square-root of the potential equation~(\ref{FcnVEqn}), namely,
\begin{gather*}
p (  f_{1} (  x,s )  ,s )  =\left(  \frac{d}{dx}f_{1} (
x,s )  \right)  p (  x,s )  . 
\end{gather*}
(In the mathematical literature, an equation of this form is known as a
``Julia functional equation''~\cite{j,k}.)
  The resulting momentum is multi-valued, in general, with the various
branches given by analogues of the recursion relation~(\ref{FcnUEqn}),
appropriately signed. That is to say,
\begin{gather*}
p_{\underset{\text{$n$ times}}{\underbrace{\pm\pm\cdots\pm}}} (  x )
=\pm\sqrt{s (  s-4x )  } p_{\underset{\text{$n-1$ times}
}{\underbrace{\pm\pm\cdots\pm}}} (  x_{\pm} )  ,
\end{gather*}
with $x_{\pm}$  as def\/ined in (\ref{xPlusMinus}). Some of these are actually
complex-valued, so for real trajectories, it is necessary to choose from among
the $2^{n}$ possibilities at each stage of the recursion. (See the
discussion in~\cite{cv}, especially for the $s=10/3$ case.) For $2<s\leq3$,
and for $s=4$, the complete inf\/inite sequence of real-valued momentum branches
is given recursively by
\begin{gather*}
p_{n} (  x )  =-\sqrt{s (  s-4x )  } p_{n-1}\left(  \frac
{1}{2}+\frac{1}{2}\sqrt{1-4x/s}\right)  .
\end{gather*}
For $3<s<4$, it is necessary to consider additional branches from among the
$p_{\pm\pm\cdots\pm}\left(  x\right)  $.

I illustrate this by plotting the f\/irst f\/ifteen branches of a zero-energy
phase-space trajectory for the marginal $s=3$ case. In some ways this is
representative of all $s>2$ cases. Among other things, the plot shows that
the system is \emph{quasi-Hamiltonian}~\cite{cz2} since the resulting
trajectory is \emph{not} single-valued. On the other hand, the trajectory
possesses such elegant features that I cannot help but make the suggestion
that the underlying analytic potential is somehow fundamental, perhaps even
expressible as a known function.

\begin{figure}[h!]\centering
\includegraphics[width=120mm]{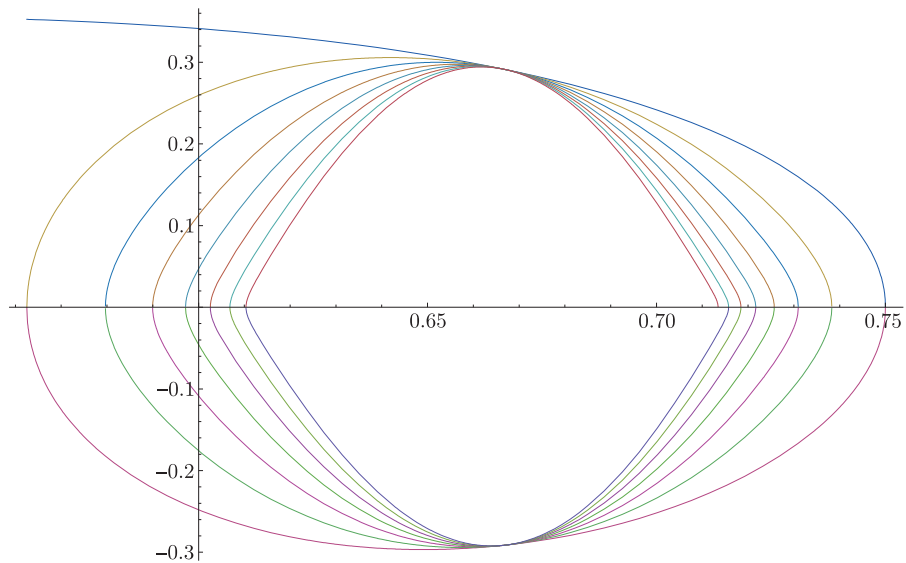}

\caption{Phase-space trajectory for $s=3$, as given by $dx/dt$ versus~$x$. As $t$
advances, the trajectory converges horizontally, but not vertically, towards
$x_{\ast}=\frac{2}{3}$.}\label{Fig11}
\end{figure}

One feature that stands out in the graph is the value of $p_{n}(
x_{\ast})  $ for $x_{\ast}=2/3$. From the recursion relation in this
$s=3$ case,
\begin{gather*}
p_{n} (  x )  =-\sqrt{3 (  3-4x )  } p_{n-1}\left(  \frac
{1}{2}+\frac{1}{2}\sqrt{1-4x/3}\right)  ,
\end{gather*}
it follows that $p_{n} (  2/3 )  =-p_{n-1} (  2/3 )  $,
exactly. Thus, while the phase-space trajectory contracts horizontally,
towards the asymptotic f\/ixed point at $x_{\ast}=2/3$, the vertical height at
$x_{\ast}=2/3$ is \emph{fixed and nonzero}. Numerically, we f\/ind $\left\vert
dx/dt\right\vert _{x=2/3}=0.291464$. Remarkably, the transit times for the
corresponding $V_{n\geq1}$ branches of the potential are all $\Delta t=1$,
even though those branches also shrink to zero width as $n\rightarrow\infty$,
but with f\/ixed depth, $V_{n}\left(  2/3\right)  =-0.0849511$.

For additional details, and numerical results, please see \cite{cz1,cz2,cv}.
  However, I do not wish to give the impression that there is nothing further
to be done here. To determine the large time behavior of maps, there is much
to be done. While the asymptotic form of $\Psi^{-1}$ can be determined for
non-chaotic values of $s$, I do not know what it is for generic, chaotic
versions of the logistic map. The one chaotic case where it is known
exactly, in simple closed-form (i.e.~$s=4$), is easily understood, but raises
many questions about the behavior for other nearby values of~$s$.

\appendix

\section[Series for the $s=1$ logistic map potential]{Series for the $\boldsymbol{s=1}$ logistic map potential}\label{appendixA}

For this case the appropriate limit of (\ref{LogisticPotentialSeries}) gives%
\begin{gather*}
V (  x,1 )  \equiv\lim\limits_{s\rightarrow1}V (  x,s )
=-x^{4}-2x^{5}-4x^{6}-\frac{25}{3}x^{7}-\frac{215}{12}x^{8}-\frac{589}%
{15}x^{9}-\frac{7813}{90}\allowbreak x^{10}\\
\phantom{V (  x,1 )  \equiv}{} -\frac{60 481}{315}x^{11}
-\frac{11 821}{28}x^{12}+O\big(  x^{13}\big)    .
\end{gather*}
More systematically, let%
\begin{gather}
V (  x,1 )  =-x^{4}\left(  1+\sum_{n=1}^{\infty}c_{n} x^{n}\right)
 ,\qquad  c_{1}=2  ,\qquad c_{2}=4  ,\qquad c_{3}=\frac{25}{3}  ,\qquad \dots ,\label{V(s=1)Series}
\end{gather}
and solve by iteration the functional equation that should be obeyed by
$V (  x,1 )  $, namely,
\begin{gather*}
V (  x (  1-x )  ,1 )  = (  1-2x )  ^{2}V (x,1 )  .
\end{gather*}
If the formal series (\ref{V(s=1)Series}) is constructed to $O\left(
x^{25}\right)  $ or so, it appears to have a radius of convergence of
$R\approx1/2$. But, this convergence is seen to be illusory to higher
orders. Dif\/ferent behavior sets in around $O\left(  x^{30}\right)  $, where
the successive $ \vert c_{n} \vert ^{1/n}$ used in the $\lim\sup$
determination of~$R$, (\ref{Radii}), begin to grow linearly, on average, for
$n>30$, as shown here (purple curve, with wiggles).

\begin{figure}[h!]\centering
\includegraphics[width=120mm]{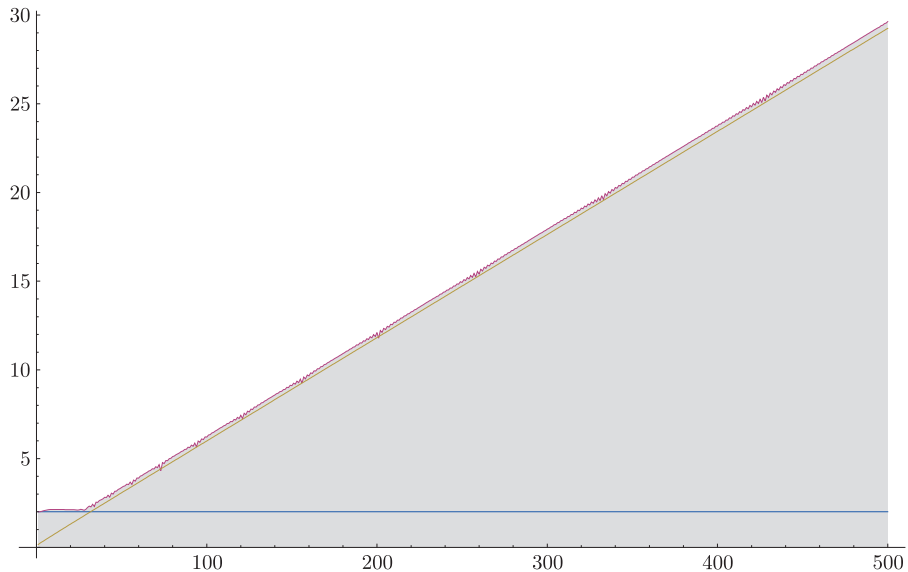}

\caption{$ \vert c_{n} \vert ^{1/n}$ for $s=1$ and $1\leq n\leq500$. The
blue horizontal is\ $2$, while the sienna line is $\left(  2^{-n/2}
e^{-3n/2}n!\right)  ^{1/n}$.}\label{Fig12}
\end{figure}

That is to say, $\left\vert c_{n}\right\vert \sim L^{n}\exp\left(  n\ln
n\right)  $ for large $n$, with $L\approx\frac{16}{270}\approx\frac{1}%
{\sqrt{2}e^{5/2}}=\allowbreak5.\,\allowbreak804\,29\times10^{-2}$. This
immediately brings to mind Stirling's formula, $n!\sim\sqrt{2\pi n}\left(
\frac{n}{e}\right)  ^{n}$, and in fact, a comparison of $c_{n}$ with
$f_{n}\equiv2^{-n/2}e^{-3n/2}n!$ is striking, as shown in  Fig.~\ref{Fig12}. (The
most important feature in the f\/igure is agreement between the averaged slopes.
 A~better overall f\/it to the~$c_{n}$, on average, is achieved by shifting the
sienna line slightly to the left, for example by including an additional
$\sqrt{n}$ factor in $f_{n}$.)

All this is compelling evidence that (\ref{V(s=1)Series}) itself is \emph{not}
convergent, but rather an asymptotic series. This is consistent with general
results in the mathematical literature \cite{e,k}. Indeed, the~$f_{n}$
``f\/it'' to $c_{n}$ can be obtained from an
asymptotic approximation of a simple integral:
\begin{gather*}
\mathcal{I} (  x )      \equiv\int_{0}^{\infty}e^{-y}\frac{1}
{1-\frac{xy}{\sqrt{2e^{3}}}}dy=-\frac{\sqrt{2}}{x} \exp\left(  \frac
{3x-2\sqrt{2}e^{\frac{3}{2}}}{2x}\right)  \operatorname{Ei}\left(  1,-\frac
{1}{x}\sqrt{2}e^{\frac{3}{2}}\right) \nonumber\\
 \phantom{\mathcal{I} (  x )}{}   =\int_{0}^{\infty}e^{-y}\sum_{n=0}^{\infty}x^{n}2^{-n/2}e^{-3n/2}
y^{n}dy\sim\sum_{n=0}^{\infty}f_{n} x^{n}  .
\end{gather*}
Taking the Cauchy principal value for $\mathcal{I} (  x )  $ gives
f\/inite numerical results for all~$x$. These results might be useful to
compute corrections to the polynomial approximation for $V (  x,1 )
$ as constructed from truncating the series (\ref{V(s=1)Series}). I leave
further analysis of this interesting but peculiar case to the reader.

\subsection*{Acknowledgements}


\looseness=-1
I would like to thank the organizers of the workshop on \emph{Supersymmetric
Quantum Mechanics and Spectral Design}, Centro de Ciencias de Benasque Pedro
Pascual, for the excellent job they did, and for giving me the opportunity to
talk about this work.  I~also thank Andrzej Veitia and Cosmas Zachos for
sharing their thoughts about functional evolution methods, and the anonymous
referees for suggestions and questions that led to improvements in the
manuscript.  Finally, I thank the CERN Theoretical Physics Group for its
gracious hospitality and generous support during my sabbatical in~2010.  This
work was also supported in part by NSF Award 0855386.

\pdfbookmark[1]{References}{ref}
\LastPageEnding

\end{document}